# Character of the reaction between molecular hydrogen and silicon dangling bond in amorphous SiO$_2$

F. Messina*, M. Cannas

Dipartimento di Scienze Fisiche ed Astronomiche - Università di Palermo. Via Archirafi 36, I-90123 Palermo, Italy.

*email: fmessina@fisica.unipa.it

TITLE RUNNING HEAD: Reaction between H$_2$ and E' center in silica.

CORRESPONDING AUTHOR FOOTNOTE: email address: fmessina@fisica.unipa.it; telephone: +390916234218; fax: +390916162461

ABSTRACT: The passivation by diffusing H$_2$ of silicon dangling bond defects (E' centers, O$_3$≡Si•) induced by laser irradiation in amorphous SiO$_2$ (silica), is investigated *in situ* at several temperatures. It is found that the reaction between E' center and H$_2$ requires an activation energy of 0.38eV, and that its kinetics is not diffusion-limited. The results are compared with previous findings on the other fundamental paramagnetic point defect in silica, the non bridging oxygen hole center, which features completely different reaction properties with H$_2$. Besides, a comparison is proposed with literature data on the reaction properties of surface E' centers, of E' centers embedded in silica films, and with theoretical calculations. In particular, the close agreement with the reaction properties of surface E' centers with H$_2$ leads to conclude that the bulk and surface E' varieties are indistinguishable from their reaction properties with molecular hydrogen.

1. INTRODUCTION

Amorphous silica (a-SiO$_2$) is a material of major scientific interest for its physical characteristics peculiar of the glassy state and for the many technological applications, particularly in optics and microelectronics. Generation of point defects, induced by laser or ionizing radiation, limits the applicative performance of the material, causing in particular the reduction of its native high optical transparency in the ultraviolet (UV).[1] The stability of point defects in a-SiO$_2$ is often conditioned by mobile species like hydrogen and oxygen, able to diffuse in the matrix in a wide temperature range.[1-5] The migration of mobile species in amorphous solids and their reactivity with point defects are timely research issues motivated by both a fundamental and an applicative interest.[4-7] In particular, diffusion of hydrogen in SiO$_2$ has been widely discussed due to the ubiquitous presence of this impurity in the material.[1-4] In this context, one of the basic processes is the reaction between H$_2$ and the silicon dangling bond (O$_3$≡Si•) defect in silica, known in the specialized literature as E' center:[1-2,8-17]

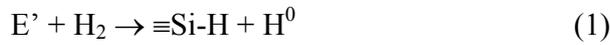

$$E' + H_2 \rightarrow \equiv Si\text{-}H + H^0 \qquad (1)$$

The E' center is a paramagnetic defect virtually found in every irradiated a-SiO$_2$ specimen, and its microscopic structure consists in an unpaired sp$^3$ electron localized on a three-fold coordinated Si atom. This defect features a wide absorption band peaked at 5.8eV which usually dominates the UV optical absorption (OA) profile of the irradiated material.[1,17] In the presence of H$_2$, reaction (1) causes the partial decay (passivation) of E' centers in a time scale of a few hours at room temperature.[8-9,15-17] This process finds a place in the general issue of passivation of many defects detrimental for applications; other important examples are the P$_b$ center, (Si$_3$≡Si•),[6,18-19] common at the Si/SiO$_2$ interface in metal-oxide-semiconductor (MOS) devices, and the oxygen dangling bond (O$_3$≡Si-O•),[2,4] which is the other fundamental paramagnetic defect in silica glass, also known as Non Bridging Oxygen Hole Center (NBOHC).

The reaction properties between $H_2$ and NBOHC centers induced by laser irradiation have been clarified by recent experimental investigations,[4-5] which conclusively established that the reaction kinetics is diffusion-limited. This finding agrees with what is commonly expected for the reactions of diffusing species with point defects in solids; indeed, in these systems the diffusion process is usually so slow to be the actual bottleneck controlling the reaction rate. In these conditions, the characteristic activation energy of the reaction coincides with that of $H_2$ diffusion in silica, $\Delta E(H_2) \sim 0.4 eV$;[2,4] consistently, it was also shown that the kinetics of the reaction between NBOHC and $H_2$ is conditioned by the site-to-site statistical distribution of $\Delta E(H_2)$ in the glass,[2] known to be a fingerprint of the disordered structure of amorphous silica.[2,4]

In contrast, much less is known about the reaction properties of E' center. In fact, although the passivation of E' center by $H_2$ has been discussed by many experimental and theoretical works, aiming to estimate the kinetic parameters of the process, it remains debated whether the reaction kinetics is diffusion- or reaction-limited.[9-16] On the one hand, several theoretical studies have proposed the reaction to require an activation energy of a few tenth of eV.[10,12-14] On the other hand, until now detailed experimental kinetic studies have been carried out only for surface E' centers,[11] and for E' centers in thin silica films,[8] and even these two works report disagreeing results: the measured rates of reaction (1) at T=300K differ by two orders of magnitude, while the activation energies are 0.4eV and 0.3eV respectively.[8,11] Most important, scanty experimental data are available for E' centers in *bulk* silica; indeed, while the passivation of E' centers on the surface (or embedded in a film) can be directly investigated by exposing the sample to gaseous $H_2$, in the case of the bulk E' center the task of studying experimentally the reaction becomes more difficult, since the concentration of available $H_2$ is no more an external controllable parameter, and the diffusion of $H_2$ in the bulk $SiO_2$ matrix becomes a necessary step to bring the reagents in contact. Moreover, it is needed to measure *in situ* the kinetics of the transient defects in temperature-controlled experiments. For these reasons, even if process (1) has been repeatedly observed to spontaneously passivate radiation-induced E' centers in bulk,[1,9,15-17] its thermal activation properties have still to be thoroughly investigated. In this paper we present an experimental study of the temperature dependence of the reaction between bulk E' and $H_2$ based on *in situ* OA spectroscopy. The main result is the estimate of the activation energy and of the pre-exponential factor of the process, which lead to the conclusion that the reaction kinetics is not diffusion-limited.

2. EXPERIMENTAL SECTION

The experiments were performed on Infrasil301 fused silica samples (provided by Heraeus Quartzglas), 5×5×2 mm$^3$ sized, OH content ~10 ppm. The samples were placed in high vacuum ($10^{-6}$ mbar) in a He flow cryostat (working between 4K and 500K), where they were UV-irradiated perpendicularly to the minor surface by the IV harmonic at 4.7eV of the pulsed radiation emitted by a Nd:YAG laser (5 ns pulse width, 1Hz repetition rate, 40mJ/cm$^2$ energy density per pulse). In order to monitor the temporal evolution of the 5.8eV absorption band of the E' centers, the OA spectrum of the sample was measured *in situ* during and after the end of each irradiation session by a single-beam optical fiber spectrophotometer, working in the 200-400nm range and equipped with a $D_2$ lamp source and a charge-coupled-device detector. Since the E' centers are paramagnetic due to the presence of an unpaired electron, they can be detected also by Electron Spin Resonance (ESR). ESR measurements on the signal of E' centers were performed at T=300K by a Bruker EMX spectrometer working at 9.8GHz, with a $8\times10^{-4}$ mW non saturating microwave power and a 0.01mT modulation amplitude. The concentration of E' centers was calculated by comparison of the signal with a reference sample with a 20% accuracy. The defects were virtually absent before irradiation, as checked by preliminary ESR measurements.

3. RESULTS AND DISCUSSION

We performed isothermal irradiations at several temperatures on different as-grown specimens. Each sample was exposed to laser radiation for a time $\tau=2\times10^3$s at a given temperature T, and the temporal evolution of its OA spectrum was monitored during exposure. Then, in order to observe the post-irradiation kinetics of the induced absorption, the measurements went on also for a few $10^3$s after the end of exposure, while keeping the specimen at the same temperature T at which it had been irradiated. After the end of OA measurements, the samples were kept some more hours at T, after which they returned to 300K where we continued to monitor the E' centers by ESR for a few days. In Fig. 1 we report the typical difference OA profile in the UV, as measured at several times during (a) and after (b) the end of the exposure session at 250K; all the curves are difference spectra calculated with respect to the OA profile of the as-grown specimen. The main induced signal is the 5.8eV band of the E' centers, which grows during irradiation (a) and decreases in the post-irradiation stage (b). The post-irradiation decay process is observed only at temperatures higher than ~200K.

Aside from the 5.8eV band, the spectra in Fig. 1 show a weak negative component at ~5eV and a broad weak component near ~4eV, both due to Ge impurities in the material,[20] and of no concern here. No other signals are detected; in particular, the NBOHC centers ($O_3\equiv Si-O\bullet$) are known to feature an absorption band peaked at 4.8eV,[1] which is absent in the measured OA profiles within 0.01cm$^{-1}$; this corresponds to a concentration upper limit of about $10^{15}$cm$^{-3}$, based on the known oscillator strength.[1,4,16]

From the peak amplitude of the 5.8eV band and using the absorption cross section of the defects,[16] we calculate the concentration [E'](t) as a function of time, reported in Fig. 2 at four representative temperatures from 200K to 350K. [E'] grows during exposure to a final value [E']($\tau$) which is a function of T.

In previous experiments at T=300K it was demonstrated that the decay of E' in a-SiO$_2$ after 4.7eV laser irradiation is due to reaction with diffusing H$_2$ of radiolytic origin. This was inferred by comparison of the kinetics with the diffusion parameters of H$_2$ and by the observation of the concurrent growth of the H(II) center (=Ge•-H).[20] In particular, it is worth noting that the presence of hydrogen in the chemical structure of the H(II) radical is unambiguously inferred by the hyperfine structure of its ESR signal, due to the interaction between the electron and proton spins;[20] hence, the H(II) center may be considered as a probe of the presence of diffusing hydrogen in the irradiated sample.[20] Moreover, Eq. (1) is basically the only active reaction, due to the absence of NBOHC, widely known to react with H$_2$,[2,4] and since the formation of H(II) involves only a minor portion of the H$_2$ population.[20] Consistently with these results, it will be reasonably assumed in the following that the reaction with H$_2$ is the only cause of E' decay observed at T>200K. We stress that this threshold is also consistent with the physical properties of H$_2$ diffusion in amorphous silica, which is known to be frozen below ~200K.[2,4]

As the main purpose of the present work is to study the reaction of E' with H$_2$, here we will focus only on the post-irradiation decay, leaving to a further paper the discussion of the generation stage of E' centers taking place during laser exposure. For this reason, we present here only the experimental results obtained in the 200K-375K temperature interval. While the lower limit is fixed by the mobility threshold of H$_2$ in silica, the upper limit is due to the progressive acceleration of the decay process with temperature; indeed, above roughly 400K the decay of E' centers was found to be too fast (completed within ~10$^2$s) to be accurately studied by the present experimental technique.

On this basis, the reaction constant k(T) between E' and H$_2$ can be estimated at each temperature using the following procedure. The variation rate of [E'] due to reaction (1), calculated within the stationary-state approximation,[2] is given by the chemical rate equation: d[E']/dt=-2k(T)[E'][H$_2$], where the factor 2 derives from two E' centers being passivated by each H$_2$.[2,17] Now, k(T) can be estimated from the inverse of the latter equation evaluated at the end of exposure (t=τ):

$$k_i(T) = \frac{d[E']}{dt}(\tau)\frac{1}{2([E'](\tau)[H_2](\tau))} = \frac{d[E']}{dt}(\tau)\frac{1}{[E'](\tau)([E'](\tau)-[E']_\infty)} \qquad (2)$$

where $[E']_\infty$ is the stationary concentration measured at long times, and the last equality derives from the fact that the total post-irradiation decrease of $[E']$, $[E'](\tau)-[E']\infty$, is twice the amount $[H_2](\tau)$ available at the end of exposure. In Eq. (2), the temperature dependence of all the quantities at the right side is implicit, and the index "i" reminds that k is estimated from the initial decay rate of E'. A similar approach recently permitted to estimate $k_i(T=300K)=8.3\times10^{-20}cm^3s^{-1}$ in wet fused silica.[17] The initial decay slope $d[E']/dt(\tau)$ immediately after the end of irradiation which appears in Eq. (2) can be estimated by a linear fit in the first ~50s of the post-irradiation stage, as shown in the inset of Fig. 2 for the kinetics at T=350K. To calculate $k_i(T)$ we still need $[E']_\infty$ for each irradiation; for instance, at T=300K we determined that $[E']$ tends to a stationary value of $[E']_\infty(T=300K)=8.8\times10^{15}cm^{-3}$, by ESR measurements a few days after the end of exposure.[21] Also for the other kinetics, $[E']_\infty$ was measured by ESR after that the sample had returned to room temperature. For irradiations at T>300K this is clearly correct, because the samples had remained many hours at T before returning to 300K, so as to be sure that the decay was completed. Actually, this is clearly correct for *all* the kinetics, as a necessary consequence of the fact that reaction (1) is the only process causing the decay of E': in fact, $[E']_\infty$ depends solely on the irradiation temperature, since it is uniquely determined by the relationship $[E'](\tau)-[E']_\infty=2[H_2](\tau)$. The effect of returning back the sample to 300K after a few hours is just the acceleration or the slowing down of the decay, if still in progress, not altering the stationary concentration. Measuring the stationary concentration after that the sample has returned to room temperature is particularly convenient for low temperature experiments, where an *in situ* measurement of $[E']_\infty$ would require to follow the slow decay kinetics for a very long time.

Hence, from $[E']_\infty$, $[E'](\tau)$ and $d[E']/dt(\tau)$ we determined $k_i$, from Eq. (2), which is reported against temperature in Fig. 3. The uncertainties were estimated by repeating many times the experiment at T=300K. The data in Fig. 3 are consistent with an Arrhenius dependence: $k_i=A_i\times exp(-E_i/k_BT)$; from a linear fit, we obtain: $E_i=(0.28\pm0.01)eV$ and $A_i=3\times10^{-15}cm^3s^{-1}$.

We must now take into account the effects of the disorder in the glass matrix, which manifests itself in an inhomogeneous distribution of the activation energy, and consequently also of the reaction constant k. In detail, as anticipated in the Introduction section, it is known that the chemical kinetics of point defects in silica can be fitted only by a linear combination of solutions of the chemical rate equations, obtained with different reaction constants $k=A\times exp(-\varepsilon/k_BT)$ weighted on a gaussian distribution of $\varepsilon$ centered at a mean value $E_m$.[2,4,20] The pre-exponential factor A is usually taken as undistributed. In a recent experiment at T=300K, we estimated the standard deviation $\sigma$ of the distribution to be $\sigma_0=(0.05\pm0.01)$eV,[20] by fitting the post-irradiation kinetics of E' centers. Now, due to the distribution the value $k_i$ estimated from the initial slope actually corresponds to the *mean* $<k>$ value calculated on the distribution of $\varepsilon$, which is $k_i=<k>=k_m\times exp(\sigma^2/2k_B^2T^2)$, where $k_m=A\times exp(-E_m/k_BT)$. This means that $k_i$ is higher than the value $k_m$ corresponding to the mean activation energy $E_m$. As a consequence, the above derived value $E_i=0.28$eV must be actually interpreted as a phenomenological *effective* energy, describing the activation of the fastest stages of the decay. The analysis based on the initial decay slope actually samples only the most reactive subset within the whole population of species participating in the reaction. Strictly speaking, only $k_m$ is expected to exhibit Arrhenius behaviour, whereas $k_i$ should deviate from it due to the term in $1/T^2$ introduced by the distribution. Likely, these deviations are not sufficiently noticeable in the investigated temperature range.

From the physical point of view, the parameter $k_m$ appears most suitable to be compared with theoretical predictions. Hence, to estimate the mean activation energy $E_m$ we analyze the temperature dependence of $k_m$, which can be calculated from $k_i$ inverting the last equation: $k_m=k_i\times exp(-\sigma^2/2k_B^2T^2)$. In the simplest assumption, in this calculation we fixed $\sigma=\sigma_0$ at all temperatures. $k_m$ (full symbols, Fig. 3) is consistent with an Arrhenius dependence from T, and the best fit values of mean activation energy and preexponential factor for the process are $E_m=(0.38\pm0.04)$eV, and $A=3\times10^{-14}$cm$^3$s$^{-1}$.

Reactions of bulk defects in solids with mobile species are generally assumed to be diffusion-limited,[2-4,10] meaning that their rate is determined by the mobility of the diffuser, whose migration is usually the bottleneck of the process. In Waite's treatment of purely diffusion-limited reactions,[22] the rate constant of Eq. (1) is given by $k_{WD}=4\pi r_0 D$, where $D=D_0 \exp(-\Delta E(H_2)/k_B T)$ is the diffusion constant of $H_2$ in silica, and $r_0$ is the so-called capture radius of the defect, i.e. a distance under which $H_2$ is assumed to react irreversibly with the E' center. In Fig. 3 we report (dashed line) $k_{WD}$ calculated with $r_0=5\times 10^{-8}$cm, and $D_0$ and $\Delta E(H_2)$ set to literature values ($D_0=5.65\times 10^{-4}$cm$^2$s$^{-1}$, $\Delta E(H_2)=0.45$eV).[2,3,23] We observe that $k_{WD}$ is everywhere much higher than the experimental data. This is consistent with the unrealistically small value of $r_0$ which was found in previous works trying to reproduce the kinetics of reaction (1) at T=300K within Waite's theory.[9,16]

For the measured macroscopic reaction rate to be so small, it must be limited also by the rate of the reaction itself, i.e. the interaction between $H_2$ and E' below the pair separation $r_0$, rather than by the diffusion of $H_2$ only. These effects have been included in the expression of the rate constant by assuming that below $r_0$ the reaction is no longer diffusion-controlled, but proceeds by first order chemical kinetics, characterized by a rate constant $w$[s$^{-1}$].[22,24] Hence, the generalized expression of the overall rate constant is found to be:

$$k_{WR} = 4\pi r_0 D \frac{w}{w + Dr_0^{-2}} \qquad (3)$$

When $w \gg Dr_0^{-2}$, $k_{WR} \approx k_{WD}$ and the reaction is diffusion-limited, but if $w \ll Dr_0^{-2}$, $k_{WR} \approx 4\pi r_0^3 w$; in this case, the overall rate $k_{WR}$ is much smaller than $k_{WD}$ and, most important, is not anymore related to the diffusion constant D. From Fig. 3 it is now clear that reaction (1) falls in the latter case: the theory of diffusion-limited reactions is inapplicable to this reaction, whose rate is mainly determined by the reaction process in itself rather than by migration of $H_2$. It is worth stressing that this conclusion is independent from the simplifications used to treat the effects of the statistical distribution of the activation energy, because it is founded only on the result: $k_m < k_i \ll k_{WD}$.

We proceed now to compare our findings with the existing literature on the subject. The present experimental results may be used as a benchmark to be compared with computational works. Indeed, several estimates of the activation energy for process (1) ranging from 0.2eV to 0.8eV were reported, depending on the calculation method,[10,12-14] thus giving rise to a debate on the most suitable technique to calculate the reaction parameters. In this sense, the closest agreement is found with the ~0.5eV values predicted by Vitiello et al,[12] and Kurtz et al.[14]

Our result on the silicon dangling bond (E' center) strikingly differs from what was found on the other basic defect in silica, the oxygen dangling bond (NBOHC center), whose reaction with $H_2$ is diffusion-limited.[2,4] This remarkable difference indicates that, differently from NBOHC, the recombination of E' with $H_2$ is not spontaneous, but requires overcoming an energy barrier, to such an extent that the character of the reaction is overwhelmingly determined by the local features of the defect rather than by the matrix in which it is embedded. The different reactivities of NBOHC and E' with $H_2$ had been anticipated by some computational works.[1,12] The comparison with NBOHC also allows to infer an important information on the physical interpretation of the statistical distribution of the activation energy. Indeed, since the passivation of NBOHC by $H_2$ is purely diffusion-limited, the statistical distribution found for the activation energy of this reaction,[2-4] has to be interpreted as a feature inherent to the diffusion process in the amorphous matrix. For what concerns the E' center, the situation is conceptually different; in fact, since in this case the reaction rate is unrelated to the diffusion coefficient D, our best estimates for the mean activation energy $E_m=(0.38\pm0.04)$eV and the pre-exponential factor $A=3\times10^{-14}$cm$^3$s$^{-1}$, as well as the width $\sigma_0$ of the distribution of $\varepsilon$, must be interpreted as features of the local reaction, not related to the diffusion of $H_2$. In particular, it is worth to note that the similarity between $E_m$ and the activation energy $\Delta E(H_2)$ for hydrogen diffusion in silica is purely accidental. In this sense, one must distinguish between two physically different effects of disorder: inhomogeneity affecting the *diffusion process* in amorphous $SiO_2$, and inhomogeneity affecting the *reaction properties* of a specific point defect. The former manifests itself in the randomization of the diffusion constant D, and in turn of any diffusion-limited reaction constant. In contrast, the latter effect is probed by a defect whose distributed reaction constant is not diffusion-limited. The physics of silica provides good examples of these two conceptually different phenomena in the two basic defects, the NBOHC and the E' center respectively.

In regard to the reaction of *surface* E' with $H_2$,[11] where diffusion is clearly not a rate limiting factor, the mean activation energy was reported to be (0.43±0.04)eV, consistent with our value $E_m$. Also the mean rate constant k at room k(T=300K)=8.2×10$^{-20}$cm$^3$s$^{-1}$, closely agrees with our estimate $k_i$(T=300K)=8.3×10$^{-20}$cm$^3$s$^{-1}$, this leading to consistent pre-exponential factors. Since the reaction properties of surface and bulk E' with $H_2$ are found to be very similar, we conclude that the surroundings of the defect do not influence its interaction with the $H_2$ molecule, and only the $O_3\equiv Si\bullet$ mojety determines the reaction dynamics. This simple scheme is far from obvious *a priori*, since the surface and bulk varieties of E' are known to differ in basic spectroscopic features, such as their OA peak (>6eV for surface centers) and ESR signal.[1] We also observe that the $P_b$ centers ($Si_3\equiv Si\bullet$) on Si/SiO$_2$ interfaces have been found to react with $H_2$ with a much higher (~1.7eV) activation energy,[18-19] this meaning that the nature of the three silicon bonds is a factor which strongly influences the reaction parameters.

The kinetics of reaction (1) was also investigated in a pioneering work by Li et al.,[8] dealing with E' centers embedded in thermal SiO$_2$ films exposed to $H_2$ in a vessel, where it was estimated an activation energy of 0.3eV and a rate constant at room temperature of ~5×10$^{-22}$cm$^3$s$^{-1}$ which is two orders of magnitude lower than our $k_i$(300K). Consequently, the authors had qualitatively drawn the same conclusion which is rigorously confirmed here, i.e. reaction (1) is not diffusion-limited. Although the comparison with our findings could lead to conclude that the reaction parameters of E' in thin films are significantly different from bulk E', these results cannot be directly compared to ours. In fact, within the experimental approach by Li et al, one must take into account also the entry and permeation processes of $H_2$ in SiO$_2$, as additional steps in the process which may condition the measure of the reaction parameters, as clearly discussed by the same authors.[8]

## 4. CONCLUSIONS

In conclusion, we studied the temperature-dependent kinetics of laser-induced silicon dangling bond defects (E' centers) in bulk a-SiO$_2$; at T>200K, the defects decay in the post-irradiation stage by reaction with H$_2$. We demonstrate that the kinetics of this passivation process is reaction-controlled rather than diffusion-limited. In this respect, the reaction properties of E' with H$_2$ contrast with those of the other fundamental paramagnetic point defect in silica, the oxygen dangling bond (NBOHC center). We estimate the mean activation energy of the reaction, E$_m$=0.38eV, thus giving experimental support to theoretical predictions. By comparison with results on surface E' centers we conclude that the reaction properties of E' center with H$_2$ are independent from its surroundings

ACKNOWLEDGMENT: We thank R. Boscaino and group for enlightening discussions, and G. Lapis, G. Napoli for assistance in cryogenic work.

REFERENCES

(1) *Defects in SiO$_2$ and Related Dielectrics: Science and Technology*; Pacchioni, G.; Skuja, L.; Griscom, D. L., Eds.; Kluwer Academic Publishers, Netherlands, 2000; pp 73-196.

(2) Griscom, D. L. *J. Non-Cryst. Solids* **1984**, *68*, 301. *J. Appl. Phys.* **1985**, *58*, 2524.

(3) Doremus, R. H. *Diffusion of Reactive Molecules in Solids and Melts*; John Wiley & Sons, New York, 2002.

(4) Kajihara, K.; Skuja L.; Hirano M.; Hosono H. *Phys. Rev. Lett*. **2002**, *89*, 135507. *Phys. Rev. B* **2006**, *74*, 094202.

(5) Kajihara, K.; Skuja, L.; Hirano, M.; Hosono, H. *Phys. Rev. Lett*. **2004**, *92*, 015504.

(6) Rashkeev, S. N.; Fleetwood, D. M.; Schrimpf, R. D.; Pantelides, S. T. *Phys. Rev. Lett*. **2001**, *87*, 165506.

(7) Godet, J.; Pasquarello, A. *Phys. Rev. Lett.* **2006**, *97*, 155901.

(8) Li, Z.; Fonash, S. J.; Poindexter, E. H.; Harmatz, M.; Rong, F.; Buchwald, W. R. *J. Non Cryst. Solids* **1990**, *126*, 173.

(9) Imai, H.; Arai, K.; Hosono, H.; Abe, Y.; Arai, T.; Imagawa, H. *Phys. Rev. B* **1991**, *44*, 4812.

(10) Edwards, A. H. *J. Non-Cryst. Solids* **1995**, *187*, 232.

(11) Radzig, V. A.; Bagratashvili, V. N.; Tsypina, S. I.; Chernov, P. V.; Rybaltovskii, O. *J. Phys. Chem.* **1995**, *99*, 6640.

(12) Vitiello, M.; Lopez, N.; Illas, F.; Pacchioni, G. *J. Phys. Chem. A* **2000**, 104, 4674.

(13) Lopez, N.; Illas, F.; Pacchioni, G. *J. Phys. Chem. B* **2000**, 104, 5471.

(14) Kurtz, H. A.; Karna, S. P. *J. Phys. Chem. A*, **2000**, 104, 4780.

(15) Smith, C. M.; Borrelli, N. F.; Araujo, R. J. *Appl. Opt.* **2000**, *39*, 5778.

(16) Cannas, M.; Costa, S.; Boscaino, R.; Gelardi, F. M. *J. Non Cryst. Solids* **2004**, *337*, 9.

(17) Messina, F.; Cannas, M. *J. Phys. Condens. Matter* **2005**, 17, 3837.

(18) Brower, K. L. *Phys. Rev. B* **1988-I**, *38*, 9657.

(19) Stesmans, A. *Appl. Phys. Lett.* **1996**, *68*, 19.

(20) Messina, F.; Cannas, M. *Phys. Rev. B* **2005**, *72*, 195212.

(21) Consistent estimates of [E']$_\infty$ are obtained by ex situ OA, though with a lower precision than ESR.

(22) Waite, T. R. *Phys. Rev.* **1957**, *107*, 463.

(23) Shelby, J. E. *J. Appl. Phys.* **1977**, 48, 3387.

(24) Kuchinsky, S. *J. Non Cryst. Solids* **2006**, *352*, 3356.

FIGURE CAPTIONS

**Figure 1.** Typical time dependence of the induced absorption profile during (a) and after the end (b) of a laser irradiation session of 2000 laser pulses performed at $T_0$=250K. Measurements in the post-irradiation stage were performed while keeping the sample at $T_0$ after turning off the laser.

**Figure 2.** Kinetics of [E'] induced by 2000 laser pulses at several temperatures. The kinetics were vertically shifted to avoid overlap; actually [E']=0 at t=0 for all of them. Inset: zoom of the kinetics at 350K showing the least-square fit procedure used to estimate the initial decay slope.

**Figure 3.** Reaction constant between E' and $H_2$, taking ($k_m$, full symbols) or not taking ($k_i$, empty symbols) into account the distribution of activation energy. Dotted line: prediction based on a purely diffusion-limited reaction model. Full lines: fits with Arrhenius equations.

**Figure 1**

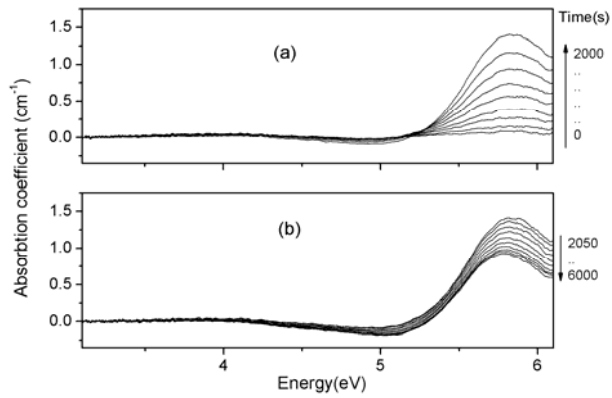

**Figure 2**

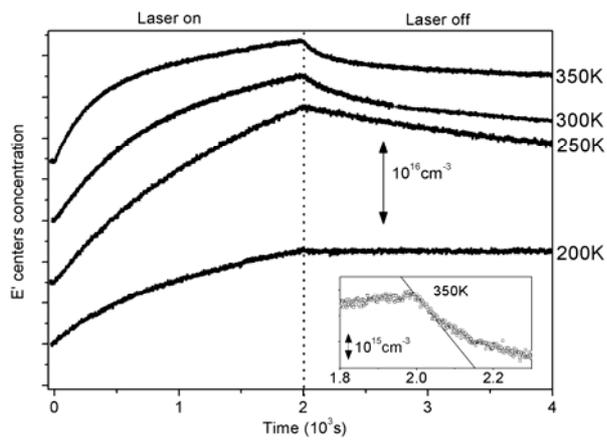

**Figure 3**

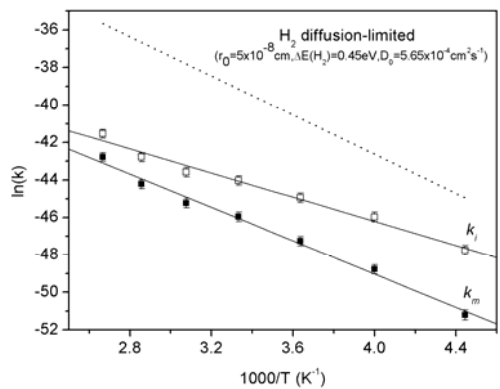

15